\documentclass[aps,prb,reprint,groupedaddress,floatfix]{revtex4-1}

\usepackage[english]{babel} 
\usepackage{amssymb,amsmath}
\usepackage{url}
\usepackage{multirow}
\usepackage{chemformula}
\usepackage{tabularx}

\usepackage{graphicx}
\usepackage{adjustbox}

\newlength{\smallpic}
\begin{document}

\title{Hundreds of new, stable, one-dimensional materials from a generative machine learning model.}
\author{Hadeel Moustafa, Peder Meisner Lyngby, Jens Jørgen Mortensen, Kristian S. Thygesen, and Karsten W. Jacobsen}
\affiliation{CAMD, Department of Physics, Technical University of Denmark, DK-2800 Kongens Lyngby, Denmark }
\date{\today}

\begin{abstract}
We use a generative neural network model to create thousands of new, one-dimensional materials. The model is trained using 508 stable one-dimensional materials from the Computational 1D Materials Database (C1DB) database. More than 500 of the new materials are shown with density functional theory calculations to be dynamically stable and with heats of formation within 0.2 eV of the convex hull of known materials. Some of the new materials could also have been obtained by chemical element substitution in the training materials, but completely new classes of materials are also produced. The band structures, electronic densities of states, work functions, effective masses, and phonon spectra of the new materials are calculated, and the data are added to C1DB.

\end{abstract}

\maketitle

\section{Introduction}

Low-dimensional materials are gaining considerable attention due to their distinctive physical and chemical properties, as well as a variety of possible applications, including light-absorbers, single-photon emitters, and catalysts \cite{spiece2019nanoscale, yang2016tuning, Cheon:2017kn}. A number of one- and two-dimensional materials have been identified and their properties have been calculated and stored in publicly available databases  \cite{lebegue_two-dimensional_2013,zhou20192dmatpedia,choudhary_high-throughput_2017,Mounet:2018ks, Cheon:2017kn, Haastrup:2018ca, moustafa2022computational}. The materials have mostly been identified based on experimental databases like the inorganic crystal structure database (ICSD) \cite{Bergerhoff:1983hj} or the crystallography open database (COD) \cite{Grazulis:2011if}, and new materials have been suggested using substitution of chemical elements with a high degree of similarity \cite{Haastrup:2018ca, moustafa2022computational}. However, the question arises whether there could be many more low-dimensional materials with interesting properties, which just wait to be discovered and synthesized.

One-dimensional materials exhibit interesting behavior and show promise for use in several fields. The reduced dimensionality results in modified band structures, charge screening, and electron-phonon coupling \cite{guo_one-dimensional_2022, balandinOnedimensionalVanWaals2022}. This opens the door to novel material properties like Luttinger liquid behavior \cite{Haldane.1981, Bruus} and the formation of Majorana bound states \cite{Wilczek.2009, majorana}. Applications could include small metallic wires \cite{stolyarov2016breakdown,xia2003one}, batteries \cite{tiwari2012zero} or transistors \cite{randle2018gate}.

Recently, a generative machine learning model was used to create new periodic materials \cite{xie2021crystal}, and the approach was subsequently adapted for two-dimensional materials\cite{lyngby_data-driven_2022}. Here we apply the same methodology to generate one-dimensional materials. We investigate their stability and classify the materials using clustering based on a distance measure between materials. Furthermore, we calculate a number of electronic properties, and provide an overview. The materials are added to the C1DB database.

The paper is organized as follows. In Section \ref{sec:methods} we describe the computational methods used for the electronic structure calculations, the machine learning model, the dimensionality classification, and the structure classification. In Section \ref{sec:Results} we describe the results including geometric classification of the new materials and a number of their properties.

\section{Computational methods}\label{sec:methods}
In the following we describe the computational techniques applied in this work. This includes electronic structure calculations, the machine learning model and the classifications of dimensionality and structure.

\subsection{Electronic structure calculations}

\subsubsection{Structure and ground state properties}
We use density functional theory (DFT) calculations as implemented in the GPAW electronic structure code \cite{Mortensen:2005ep,Enkovaara:2010jd} based on the projector augmented wave method \cite{blochl1994projector}. We apply the Atomic Simulation Environment (ASE) \cite{Larsen:2017hn} for setting up the one-dimensional structures. For our high-throughput workflow we make use of the Atomic Simulation Recipes (ASR) \cite{Gjerding.2021}, which offers a straightforward and modular framework for creating Python workflow scripts, and its automated caching mechanism monitors data provenance and keeps track of task status, which are used in conjunction with MyQueue \cite{MyQueue, mortensen2020myqueue} to control the workflow. For all calculations we use the PBE xc-functional \cite{Perdew:1996ug}, and a plane wave basis set with a cutoff energy of 800 eV.

All one-dimensional components are placed with the $z$-axis in the direction of the 1D component and a square unit cell in the $xy$-plane. It is ensured that at least 16 \AA\ of vacuum separates components in neighbouring cells. The atomic structures are relaxed using a Monkhorst-Pack k-point grid with a sampling density of 6.0 \AA, and with a Fermi temperature for smearing the electronic occupancy numbers set to 0.05 eV. The relaxation is stopped when the maximum force and maximum stress are less than 0.01 eV/\AA\ and 0.002 eV/$\textrm{\AA}^3$, respectively. For subsequent calculations of ground state properties, the k-point sampling is increased to a density of 12.0 {\AA}.

We use the elemental, binary, and ternary materials from the OQMD \cite{Kirklin:2015cr} database as a reference to calculate the convex hull for the one-dimensional materials. The energies of the materials are re-calculated with the GPAW code with the same settings as for the other total energy calculations in this work in order to ensure consistency.

\subsubsection{Phonons and dynamical stability}
We calculate the phonon frequencies at the $\Gamma$ and X point of the reciprocal unit cell. We do this using the frozen phonon technique in a unit cell, which is doubled along the direction of the 1D structure. The frequencies are calculated by diagonalizing the mass-weighted force constant matrix, which is obtained using finite displacements of the atoms by 0.01 {\AA}. The primary goal of the phonon computations is to determine whether imaginary frequencies, which correspond to negative eigenvalues for the dynamical matrix, are present. In such cases the obtained structure is not a minimum-energy structure and therefore dynamically unstable.

\subsubsection{Band structures}
The band structures of the one-dimensional materials are calculated from the Kohn-Sham eigenvalues using PBE and non-selfconsistent HSE06 \cite{heyd2003hybrid}. The reciprocal space is one-dimensional because we use large supercells with negligible interaction between neighboring one-dimensional components. The band structures are calculated using the density obtained in the ground state calculation and using 400 k-points in the first Brillouin zone. The calculations includes non-self consistent spin-orbit coupling. 

\subsubsection{Effective masses}
The effective masses of electrons and holes are calculated for bands that are within 100 meV of the valence band maximum and conduction band minimum, respectively, for all materials having a finite band gap. By fitting parabolas to the band structure adjacent to the VBM and CBM, the effective masses of the one-dimensional components are computed. The possibility of several bands, which might cross each other near to the extrema, constitute a hurdle for the fitting procedure. We distinguish between the various bands by defining for each electronic state at a certain $k$-point and energy a "fingerprint" made up of the states' projections onto the PAW projectors. The electronic states at various $k$-points are then connected into bands using that the fingerprint should change gradually throughout a band. As a result, nearby states get fingerprints that are as close as possible. 

\subsection{Generative machine learning model}
The generative model of choice is the recently developed Crystal Diffusion Variational AutoEncoder\cite{xie2021crystal} (CDVAE) which is an variational autoencoder for generating stable periodic materials. The model is only trained on stable materials from which it learns the data distribution of the given stable materials and, thus, the generated materials are biased towards stability. CDVAE omits the need of intermediate states like descriptors or fingerprints, which has proved difficult to use for generative models for periodic materials \cite{noh2020machine}, and instead works directly on the atomic coordinates in the generation process by using a diffusion model as the decoder. After training the CDVAE, new materials are generated by sampling from the latent space from which a neural network predicts a unit cell, the number of atoms and the composition. The predicted atoms are then randomly initialized in the predicted unit cell and the diffusion decoder then gradually unscrambles the random structure into a stable structure according to the model.
Equivariant graph neural networks are used for both the encoder and decoder, which ensures that the model is invariant to rotations and translations. 

CDVAE is designed to create crystals that are periodic in all three dimensions, however our goal is to generate 1D materials which are only periodic in one direction. In analogy to Ref.~\onlinecite{lyngby_data-driven_2022}, where the method is applied to two-dimensional materials, we fix this by introducing an artificial periodicity in the two non-periodic directions, but with a periodicity length scale that is much larger than the length scale along the 1D materials. Thus the graph neural networks of CDVAE only connects atoms within the 1D structure and it learns to create 1D structures.

We use the same hyperparameters for the model as used for the MP-20 dataset in Ref.~\onlinecite{xie2021crystal}, except that we increase the cutoff radius from 7 Å to 14 Å for the graph network in the decoder. This is to account for the reduced amount of neighboring atoms in a 1D structure compared to a 3D structure.

\subsection{Dimensionality classification}
We express the dimensionality of a material using the approach of Ref.~\onlinecite{Larsen:2019cf}, where a scoring parameter, $s_X$ is calculated for each dimensionality $X$, where $X$ can be for example 1D, 2D or 3D, but also mixed dimensionalities like a combination of zero- and one-dimensional components (denoted 01D). The classification is aimed at identifying covalently bonded materials components within a material with possible weak bonding between components. It is based exclusively on bond distances. A bond between two atoms $i$ and $j$ is defined to exist if
\begin{equation}
    d_{ij} < k(r_i^\text{cov}+r_j^\text{cov}),\label{eq:bonding}
\end{equation} 
where $d_{ij}$ is the distance between the two atoms, $r_i^\text{cov}$ and $r_j^\text{cov}$ are the two covalent radii, and $k$ is a continuous parameter. For small values of $k$ no bonds exist and the material consists of single, isolated atoms, \emph{i.e.} it is zero-dimensional, while for very large values of $k$ all atoms are connected by bonds, and the material is three-dimensional. For a given dimension $X$, an interval $[k_X^{min},k_X^{max}]$ may exist in which the material has the dimensionality $X$. This allows for the definition of a scoring parameter, $s_X$,  as $s_X = f(k_X^{max})-f(k_X^{min})$, where $f\left(x\right) =  c \cdot \max(0, x-1)^2 /(1 + c \cdot \max(0, x-1)^2)$ with $c = 1 / 0.15^2$. The scoring parameters are between 0 and 1 for all dimensionalities, and they sum up to one. A high value of $s_X$ indicates a high degree of likelihood that the material has the dimension $X$, \emph{i.e.}\ that the $X$-dimensional components are only weakly bound to each other. The reader is referred to Ref.~\onlinecite{Larsen:2019cf} for further details on the approach.

In this study we shall consider isolated one-dimensional components. For convenience they are described within supercells with large separation between the different components and they cannot form two- or three-dimensional materials. The dimensionality analysis is therefore used here only to evaluate whether a potentially one-dimensional system is in fact separated into zero-dimensional clusters or a combination of zero- and one-dimensional components.

\subsection{Geometric classification}
We use a clustering algorithm based on the root-mean-square-distance (RMSD) to geometrically classify the materials similar to the analysis in Ref.~\onlinecite{moustafa2022computational}. The RMSD between two different structures with atomic coordinates $\vec{R}_i$, $i=1,2,\ldots N$ and $\vec{R}^\prime_i$, $i=1,2,\ldots N$
is defined as
\begin{equation}
    \textrm{RMSD} = \sqrt{\frac{1}{N} \sum_{i=1}^N |\vec{R}_i-\vec{R}^\prime_i|^2},\label{eq:RMSD}
\end{equation}
where the ordering of the atoms in the two systems and their relative translation and rotation are chosen so as to minimize the RMSD. The chemical identity of the atoms is ignored. The application of RMSD to the one-dimensional components involves a few further considerations. Firstly, we merely take into account the atomic positions and disregard the atoms' chemical identities. Secondly, in the event that two distinct 1D components have differing atom counts, the systems are repeated along the $z$-direction to ensure that the two systems have the same number of atoms. Thirdly, to achieve the same length, the two unit cells are subsequently scaled in the $z$-direction. The coordinates in the perpendicular directions are not scaled differently.

The RMSD is used as the distance measure in single-linkage clustering and graphically expressed in dendrograms. Introducing a cutoff distance leads to an equivalence relation: two materials are related if they can be connected by a chain of materials, which all have distances smaller than the cutoff distance to their neighbors in the chain. The resulting equivalence classes constitute the clustering of the materials.

\section{Results}\label{sec:Results}

\begin{figure*}
    \centering
    \includegraphics[width=\textwidth]{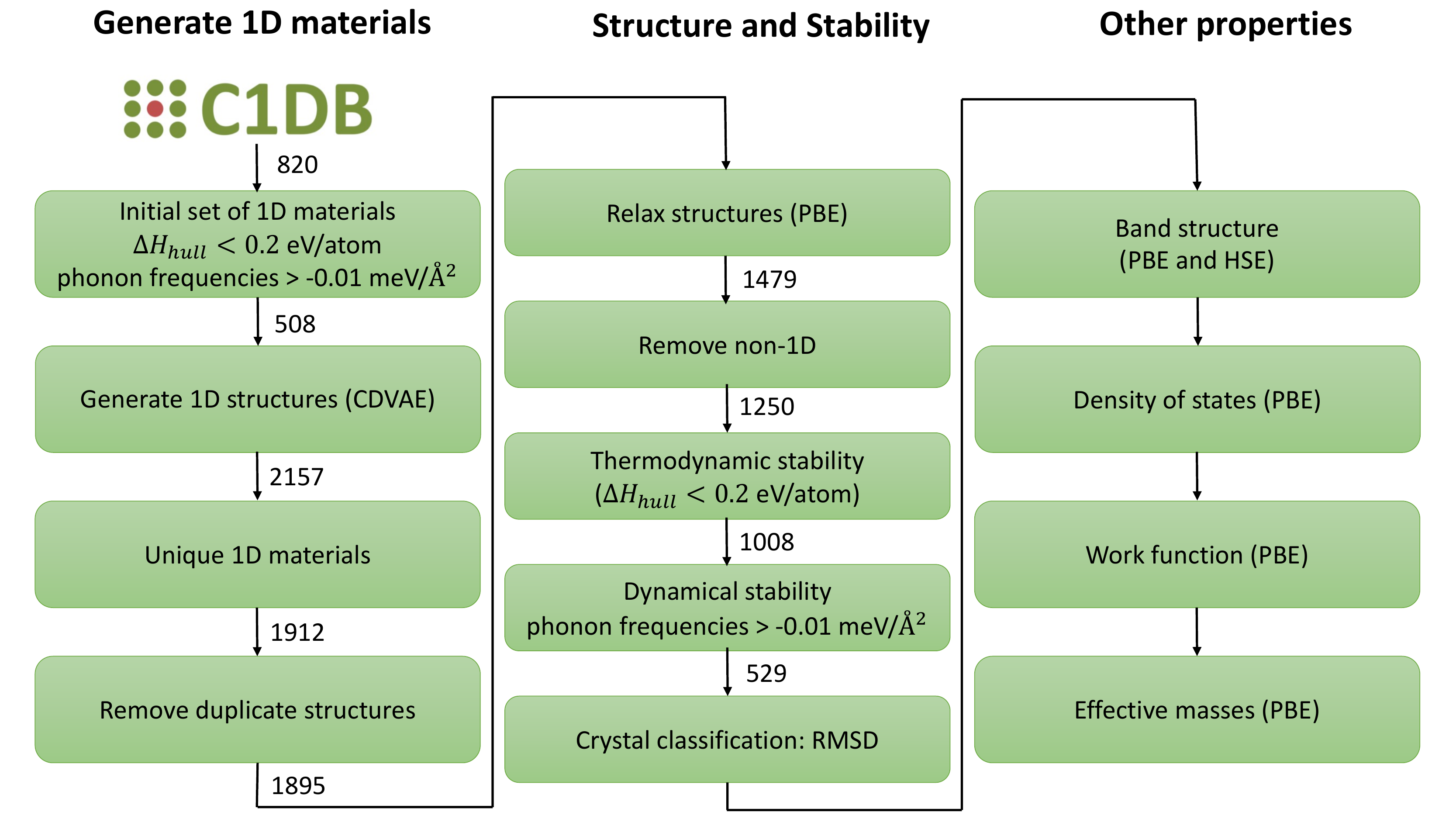}
    \caption{The figure shows the computational workflow applied in this work. The left column shows the steps involved in the training of the CDVAE and the generation and filtering of the new materials. The center column shows the steps applied in the analysis of the stability of the new materials and their geometric classification. Finally, the right column shows the additional properties calculated and stored in the updated C1DB database. The numbers indicate the number of materials considered in each step.
    The different steps are described in more detail in the text.}
    \label{fig:workflow}
\end{figure*}

Fig.~\ref{fig:workflow} depicts our computational workflow. The workflow contains both the generation of new materials, their selection, and the calculation of their properties. The workflow consists of three sub-workflows indicated by the three columns in the figure. The first sub-workflow (left column in Fig.~\ref{fig:workflow} contains the generation of new materials using the CDVAE and an initial filtering to remove materials, which are unphysical or which are clearly not one-dimensional. The second sub-workflow (middle column in Fig.~\ref{fig:workflow}) involves the study of the stability of the new materials and their structural classification, and finally the third sub-workflow (right column in Fig.~\ref{fig:workflow}) consists of the calculation of a number of electronic properties of the materials to be included in the C1DB database.

\subsection{Materials generation}

The CDVAE is trained on 508 one-dimensional materials from the C1DB database \cite{moustafa2022computational,c1db}, which are found to be stable. The thermodynamic stability of the training materials are evaluated by requiring the heat of formation to be less than 0.2 eV/atom above the convex hull. Furthermore, the dynamical stability is evaluated by considering the phonon calculations at $\Gamma$ and X. All eigenvalues of the mass-weighted force-constant matrix are required to be larger than -0.01 meV/Å$^2$.
The small negative value is chosen to remove the translation modes and to allow for small numerical errors. These criteria are also the ones used in C1DB to signify a high degree of stability.

After training, the CDVAE is used to generate 10.000 structures out of which 2003 fails the validity check of CDVAE, which checks that the bonds are above 0.5 Å. Moreover, the validity checker requires that a charge neutral combination of the elements can be formed based on the possible oxidation states of the elements, except if all the elements are metallic in which case the material is always accepted. Afterwards, the generated materials are checked for duplicate structures based on Pymatgens StructureMatcher\cite{ong2013pymatgen}, and we find that a large fraction of the generated structures are duplicate structures where out of the remaining 7997 structures only 2157 are unique. We believe that one reason for the large fraction of duplicate structures is the small size of our training set.

The resulting 2157 materials are analysed using the scoring parameter approach of Ref.~\onlinecite{Larsen:2019cf} as implemented in the ase.geometry.dimensionality module of ASE \cite{Larsen:2017hn}. Materials, which consist of one-dimensional components are selected, while materials with only zero-dimensional components are removed. If both one- and zero-dimensional components are present, the one-dimensional components are extracted and also considered in the following steps. This leaves in total 1912 materials for further analysis.
As a final step in the initial filtering we remove the materials, which are duplicates of materials already in C1DB, but not included in the training set, because they are not considered sufficiently stable. There are only 17 materials in this class, so the final number of materials to be considered further is 1895.

\subsection{Structure optimization}

The first step in the analysis of the generated materials is to perform a relaxation of the atomic coordinates and the length of the unit cell to minimize the electronic ground state energy using DFT. A significant number of the calculations (416) fail, mainly due to lack of convergence of the Kohn-Sham SCF cycle in the initial structure, which might include unusual bond lengths. All calculations are spin-polarized and the convergence problems are in many cases probably related to the determination of the magnetic structure of the materials. We discard these non-converged calculations and focus on the 1479 well-optimized materials.

\begin{figure}[htb]
    \centering
    \includegraphics[width=0.8\linewidth]{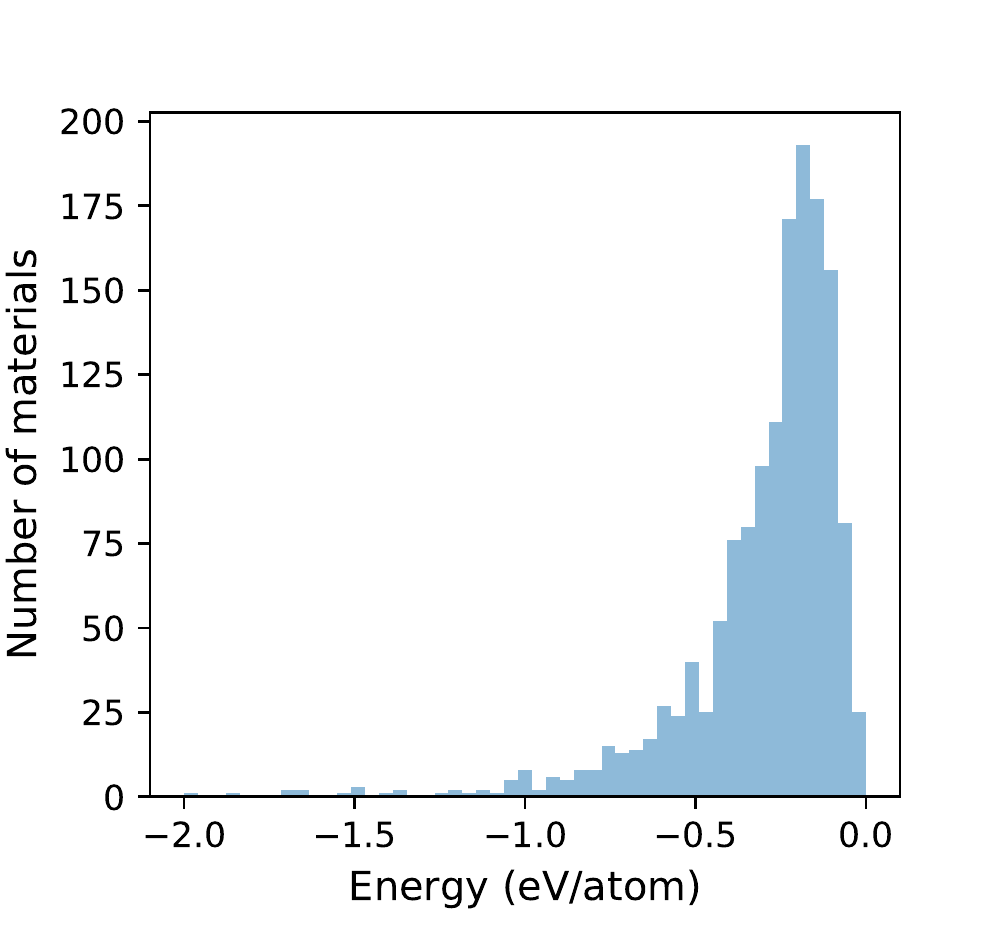}
    \caption{Histogram of the energy difference per atom between the initial and final configurations of the successfully relaxed materials.}
    \label{fig:energy_per_atom}
\end{figure}

Fig. \ref{fig:energy_per_atom} shows how much the energy (per atom) decreases during the structural relaxation. A typical relaxation energy is in the range 0.1-0.3 eV indicating that the machine-generated initial structures are in fact quite reasonable. However, in some cases a more dramatic restructuring is taking place. There are 22 materials with a relaxation energy lower than -2 eV/atom, which are outside the boundaries of the figure. One example is \ch{TiZr2Cl7} with an initial structure where the Ti and Zr atoms are very close, resulting in a huge relaxation energy of -39.5 eV/atom.

\subsection{Dimensionality}
\begin{figure}[htb]
    \centering
\begin{tabular}{ | c c c |}  \hline
Before & & After \\ \hline
\includegraphics[width=30mm]{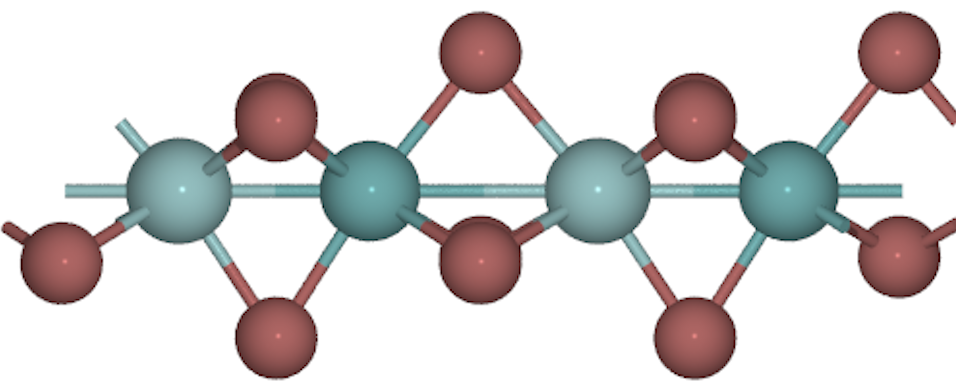} &\includegraphics[width=10mm]{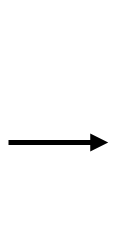}& \includegraphics[width=30mm]{fig3_Br12Mo2Ru2_after.png}\\

$s_{1}$=1, $s_{0}$=0, $s_{01}$=0   &&  $s_{1}$=1, $s_{0}$=0, $s_{01}$=0 \\  \hline

 \includegraphics[width=30mm]{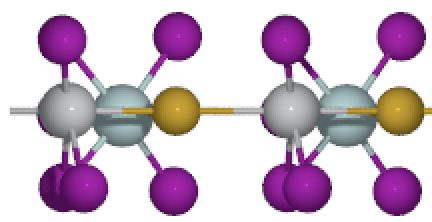} &\includegraphics[width=10mm]{to.png}& \includegraphics[width=30mm]{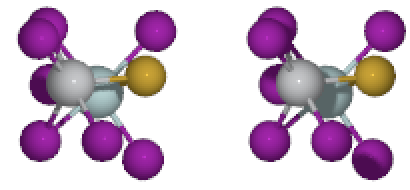}\\
 
 $s_{1}$=0.68, $s_{0}$=0.32, $s_{01}$=0   &&  $s_{1}$=0.06, $s_{0}$=0.94, $s_{01}$=0 \\  \hline

  \includegraphics[width=30mm]{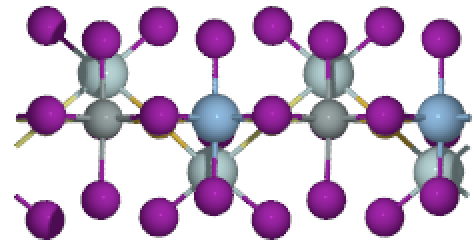} &\includegraphics[width=10mm]{to.png}& \includegraphics[width=30mm]{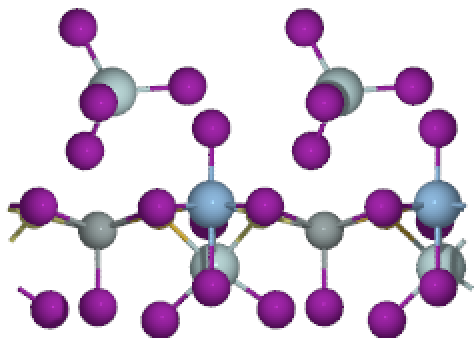}\\
  
  $s_{1}$=1, $s_{0}$=0, $s_{01}$=0   &&  $s_{1}$=0.06, $s_{0}$=0.27, $s_{01}$=0.67 \\  \hline
\end{tabular}
\caption{Examples of atomic structures before (left) and after (right) relaxation together with the calculated scoring parameters. The upper panel is showing \ch{MoRuBr6}, which remains one-dimensional during relaxation. The middle panel shows \ch{TeTiZrI7}, which decomposes into molecular, zero-dimensional components. The lower panel shows \mbox{\ch{SSeSnTaZr2I12}}, which after relaxation exhibits both zero- and one-dimensional components. Only the upper class of materials is included in the further workflow.}\label{fig:dimensionality}
\end{figure}

After relaxation we reconsider the dimensionality of the materials based on the scoring parameters. 1250 materials remain 1D in the sense that the scoring parameter $s_1$ is the largest one. 174 materials disintegrate into zero-dimensional components, while 55 materials disintegrate into both zero- and one-dimensional components. In the workflow we discard the two latter categories.  Fig.~\ref{fig:dimensionality} shows examples of structures before and after relaxation together with the associated scoring parameters $s_1$, $s_0$, and $s_{01}$.

\subsection{Thermodynamic stability and convex hull}

\begin{figure}
    \centering
    \includegraphics[width=0.8\linewidth]{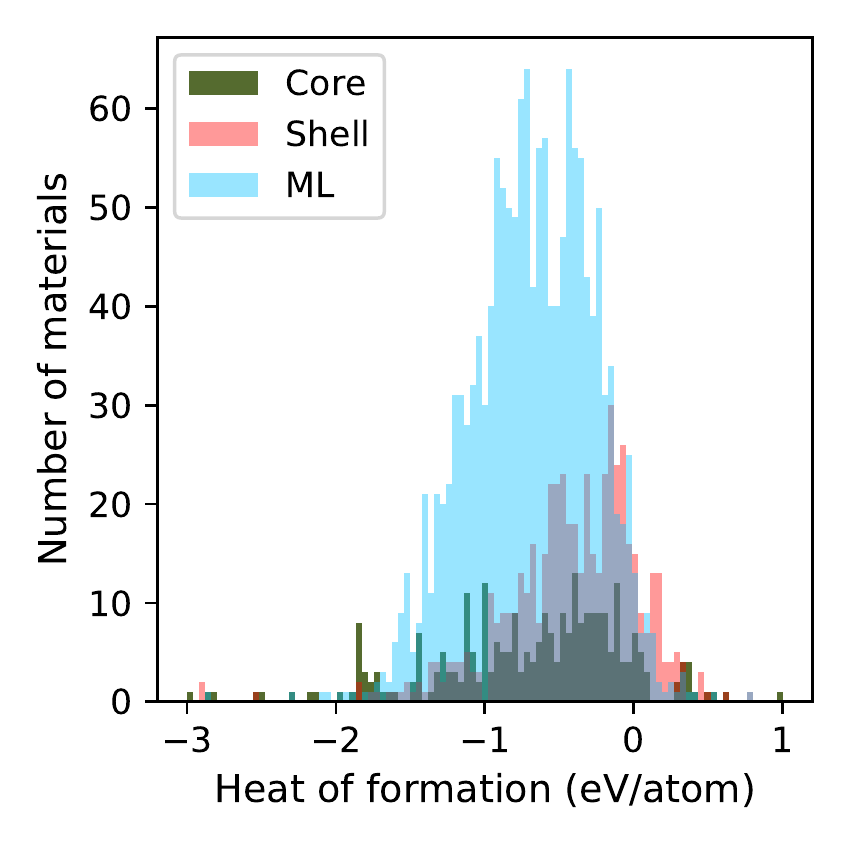}
    \caption{Distribution of the heats of formation for all the machine learning structures, and for the core and shell materials of C1DB.}
    \label{fig:heats}
\end{figure}

Fig.~\ref{fig:heats} displays a histogram of the heat of formation relative to the elements in their standard states for the machine-generated materials. For comparison, we also show the result from two classes of materials from the C1DB database: the core and shell materials. The \emph{core} materials are extracted from the inorganic crystal structure database (ICSD) \cite{Bergerhoff:1983hj} or the crystallography open database (COD) \cite{Grazulis:2011if}, while the \emph{shell} materials are obtained by substitution of chemically similar elements in the core materials. As can be seen from Fig.~\ref{fig:heats} the machine-generated materials have heats of formations quite comparable to the core and shell materials. The average heat of formation of the machine-generated materials is -0.68 eV/atom, while it is -0.70 eV/atom and -0.42 eV/atom for the core and shell materials, respectively.

\begin{figure}
    \centering
    \includegraphics[width=0.8\linewidth]{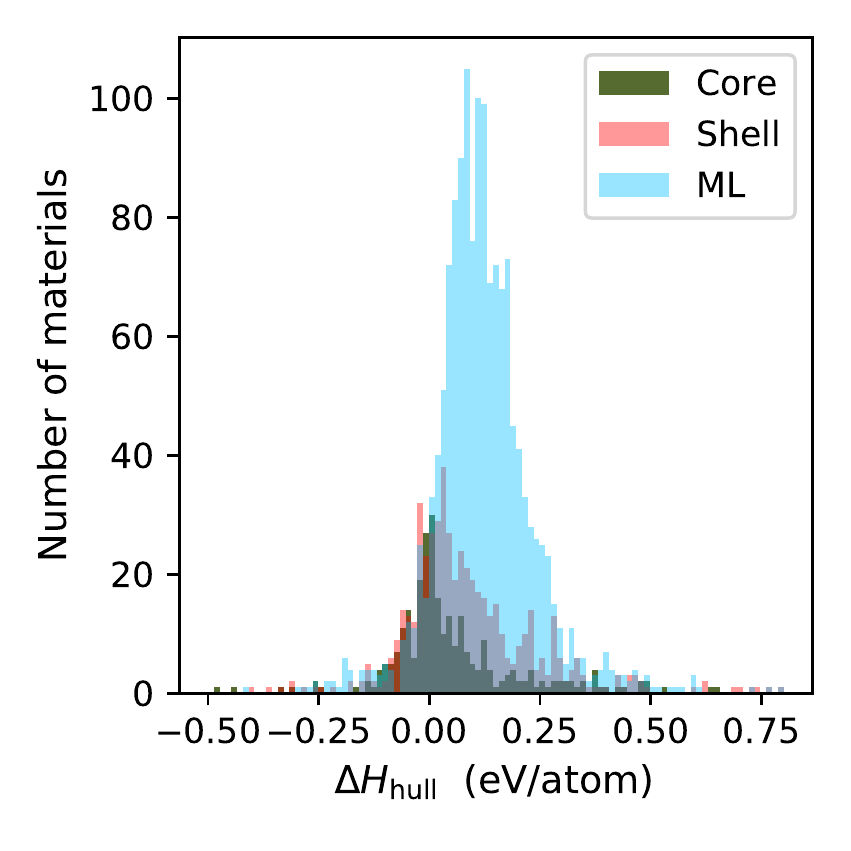}
    \caption{Distribution of the energy per atom above the convex hull for all the machine learning structures, and for the core and shell materials.}
    \label{fig:hull}
\end{figure}

The stability of the materials relative to the convex hull are illustrated in Fig.~\ref{fig:hull}, where the distribution of energies per atom above the convex hull are shown. The above-the-hull energies for the machine-generated materials are seen to be typically in the range -0.1-0.3 eV/atom, with an average value of 0.13 eV/atom. This is not much higher than for the core and shell materials, which have average values of 0.06 eV/atom and 0.09 eV/atom, respectively. In C2DB and C1DB the materials with energies relative to the convex hull below a value of 0.2 eV/atom are classified as potentially (meta-)stable, because the error in PBE is typically of that order of magnitude. Furthermore, two-dimensional materials, which can be exfoliated typically have an energy for the monolayer of the order 0.2 eV/atom or less above the convex hull \cite{Haastrup:2018ca}. Using the same convention of 0.2 eV/atom here, we see that a large fraction of the machine-generated materials (81\% or 1008 out of 1250 materials) may in fact be considered stable or meta-stable.

\begin{figure}
    \centering
     \includegraphics[width=0.8\linewidth]{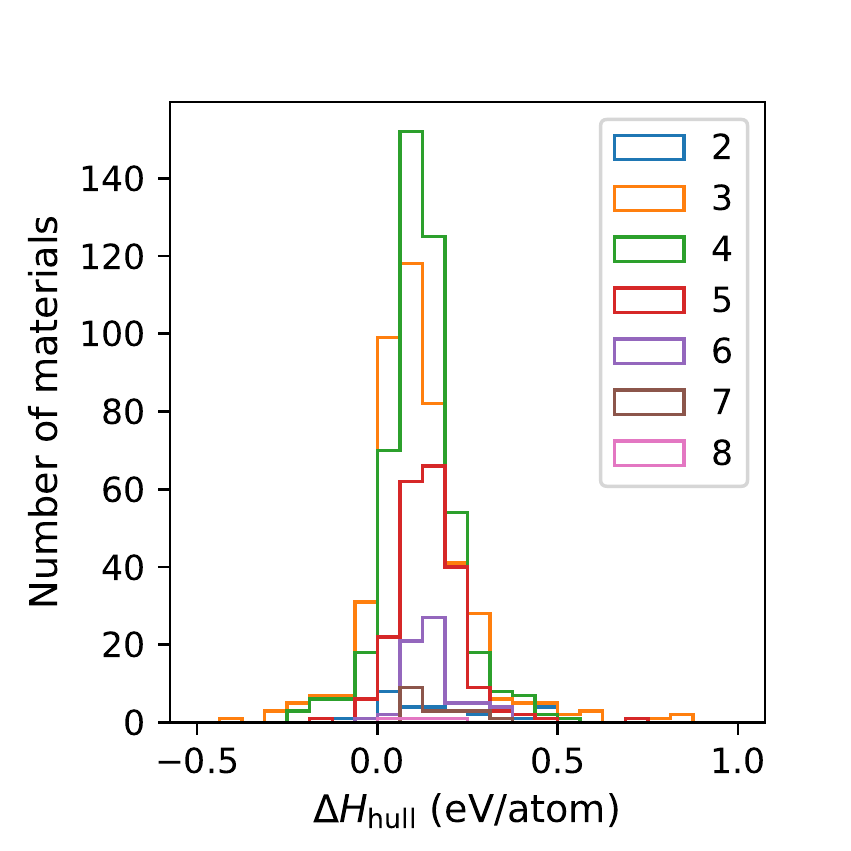}
    \caption{The distribution of the energies above the convex hull is shown for the machine-generated structures. Each distribution corresponds to a given number of different chemical elements.}
    \label{fig:elements}
\end{figure}

The CDVAE generates materials with varying number of atoms and different number of chemical elements. The training set only contains materials with up to four different chemical elements, but materials with up to eight different elements are generated. The distribution of the energy above the convex hull for the materials with different number of chemical elements can be seen in Fig.~\ref{fig:elements}. It is seen that a large fraction of the materials have three (35.9\%) or four (37.6\%) different chemical elements. The distribution of the above-the-hull-energies are rather similar independent on the number of different chemical elements.

\subsection{Dynamical stability - phonons}

\begin{figure}
\centering
\includegraphics[width=0.8\linewidth]{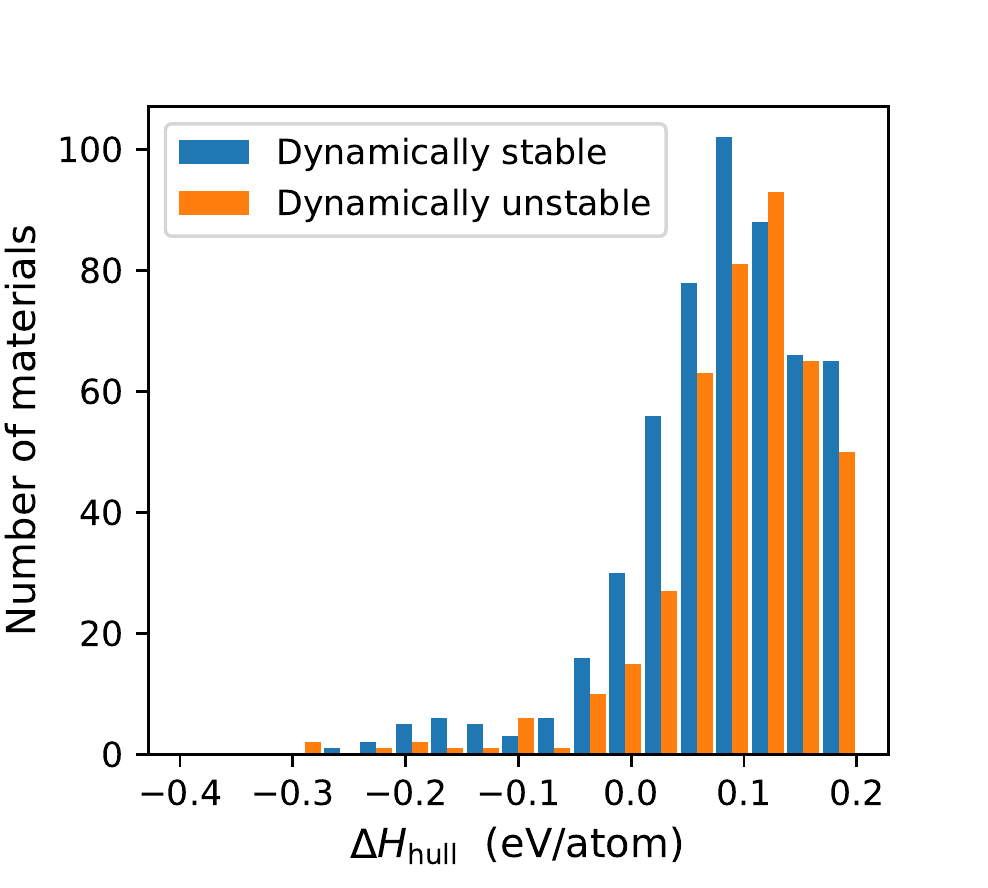}
\caption{The energy above convex hull distribution and dynamic stability 
}
\label{fig:hull_and_phonon}
\end{figure}

As a further characterization of the stability of the materials, we calculate the phonon frequencies at $\Gamma$ and X for the materials with energies above the convex hull less than 0.2 eV/atom. Imaginary phonon frequencies (\emph{i.e.} negative eigenvalues of the dynamical matrix) are an indication that the material will restructure and therefore not be stable in the form obtained in the original unit cell. To allow for the three zero-frequency translational modes and some numerical uncertainty, we consider a material dynamically stable if we do not find modes with eigenvalues less than -0.01 meV/\AA$^2$. About half of the materials (529 out of 1008) survive this criterion. As can be seen from Fig.~\ref{fig:hull_and_phonon} there is a tendency for the more thermodynamically stable materials to also be dynamically stable, but the correlation is not very strong.  

So to sum up at this stage of the workflow, the 10000 materials originally generated by the CDVAE leads to 529 new one-dimensional materials with energy above the convex hull below 0.2 eV/atom and an apparent dynamical stability at least in the doubled unit cell. We now proceed by characterizing the geometries of the new materials.

\subsection{Geometric classification}

\begin{figure}
\centering
\includegraphics[width=0.8\linewidth]{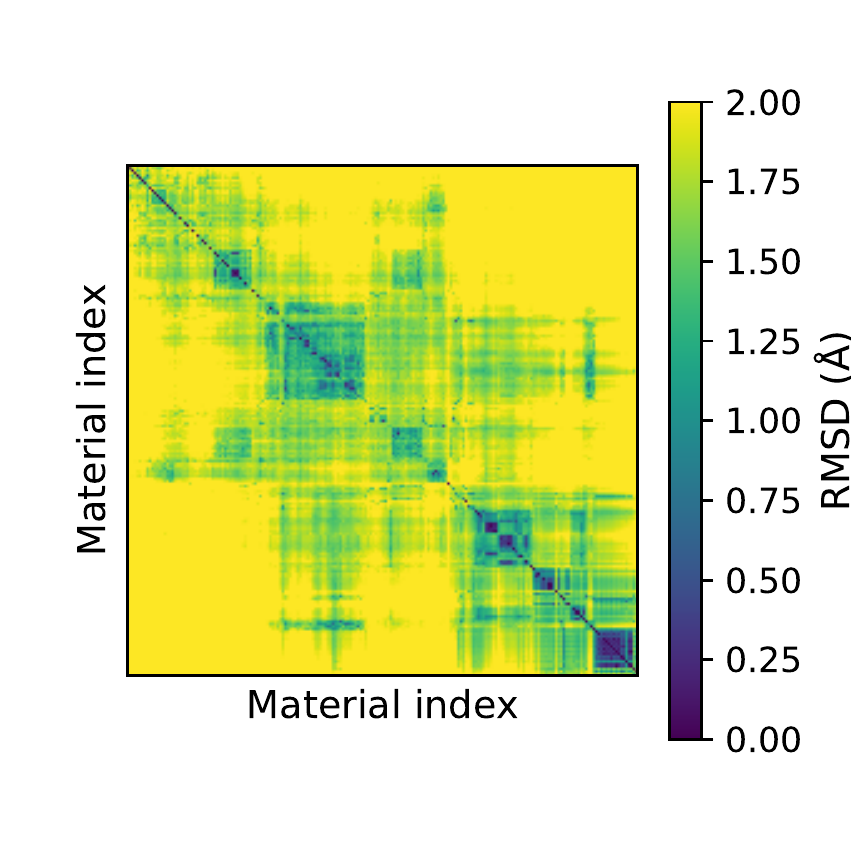}
\caption{The database's stable 1D materials' structural distance matrix. The dark squares on the diagonal represent groups of similar materials since the constructions have been arranged using the rearrangement clustering approach \cite{Zhang:2006wr}.}
\label{fig:RMSD}
\end{figure}
Fig.~\ref{fig:RMSD} shows the RMSD distance matrix between all of the 529 materials that are classified as 1D components in the database and that are stable. Utilizing the rearrangement clustering technique \cite{Zhang:2006wr}, the materials have been  permuted so that clusters with comparable structural characteristics look dark along the diagonal.

\begin{figure}
\centering
\includegraphics[width=0.8\linewidth]{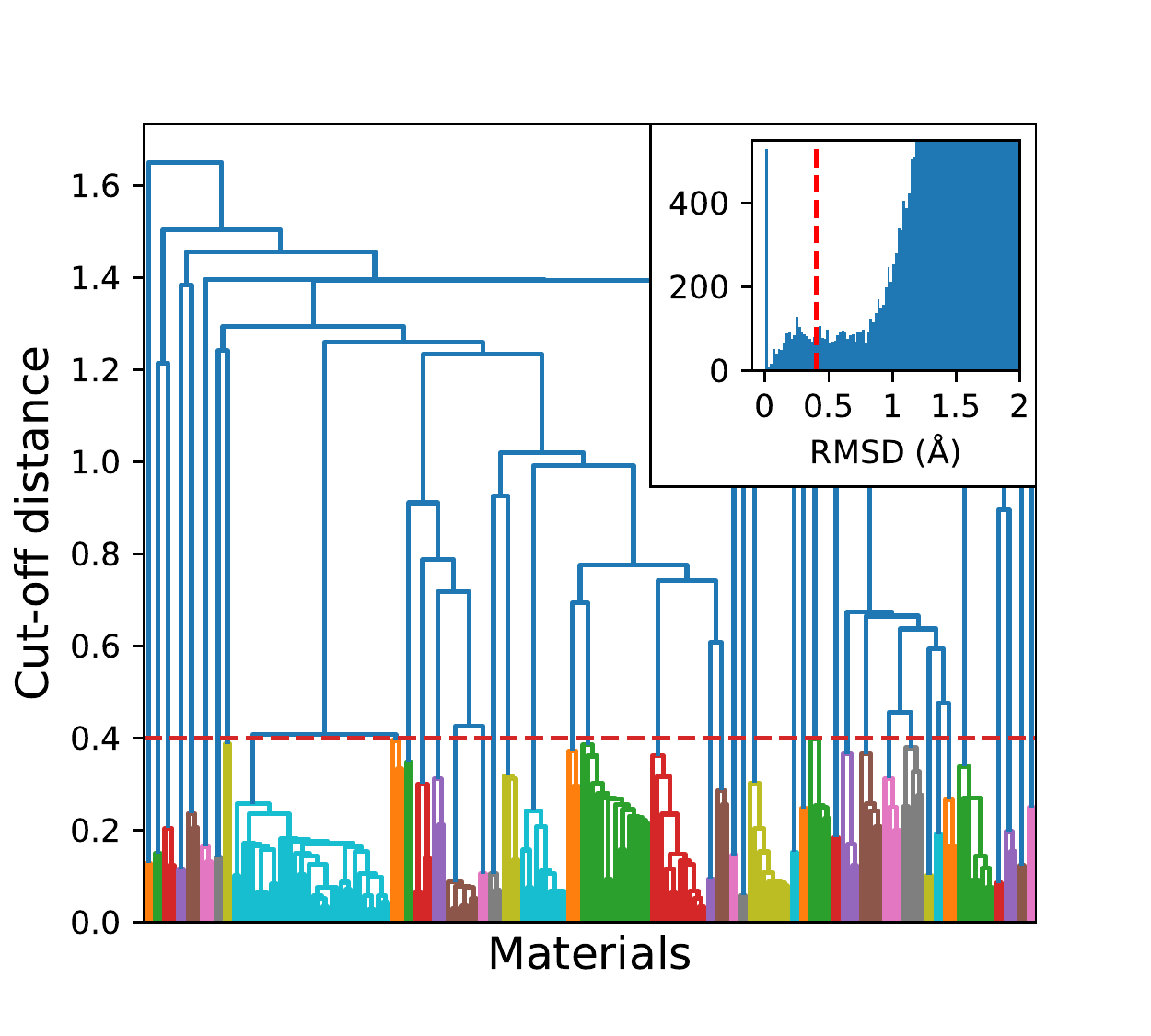}
\caption{Dendrogram showing the single-linkage clusters based on the RMSD metric. A histogram of all distances in the distance matrix is displayed in the inset, which also shows the cutoff distance of 0.4 $\textrm{\AA}$ with a vertical line. The clusters identified with a cutoff distance of 0.4 $\textrm{\AA}$ are displayed in different colors in the dendrogram.}
\label{fig:dendrogram}
\end{figure}

We now proceed to classify the new materials geometrically using RMSD and single-linkage clustering. Fig.~\ref{fig:RMSD} visualizes the distance matrix, which is obtained by calculating the RMSD between all the 529 stable materials, as a heat map. The materials have been reorganized to reveal the existence of clear clusters in the space of materials. We investigate the clustering further by constructing a dendrogram as shown in Fig.~\ref{fig:dendrogram}. The inset in the figure shows a histogram of all distances in the distance matrix.
The presence of a double-peak distribution suggests that it makes sense to separate the distances into two groups, those that represent materials, which are close by,  and those that are further apart. In a similar analysis of the core materials in C1DB, the double-peak structure was very pronounced with a clear minimum in the distribution at a distance of about $R_0 = 0.7$ $ \textrm{\AA}$ \cite{moustafa2022computational}. In the present case the separation is less clear, but we choose the distance of $R_0 = 0.4$ $ \textrm{\AA}$ to be used as a cutoff in the dendrogram. With this cutoff we identify 42 different classes of materials.

 \begin{table}
    \centering
    \begin{tabular}{|p{0.65\columnwidth}|p{0.18\columnwidth}|p{0.18\columnwidth}|}
    \hline
 Structure formula&  z direction&  y direction\\ \hline
\textbf{Group 1}:\ch{HfI3S3Ti}, \ch{Br6RuTi}, \ch{BrClI4Ti2}, \ch{I5STaTi}, \ch{Br6MoRu}, \ch{Br6NbV}, \ch{I6MoZr}, \ch{BrCl5HfMo}, \ch{I6VZr}, \ch{BrHf2I5}, \ch{I4Pd2Se2}, \ch{HfI4S2Ti}, \ch{HfI4Se2Ti}, \ch{I4Se2Zr2}, \ch{Cl6PdRe}, \ch{GeI6Ru}, \ch{GeI6Ti}, \ch{BeBr6Pt}, \ch{Bi2I6}, \ch{Bi2Br6}, \ch{GeI5TaTe}, \ch{Cl4S2Ti2}, \ch{Br3ISe2Ti2}, \ch{Cl4O2Ti2},  \ch{Br6SiTi}, \ch{I6SiTi}, \ch{Br6HfPt}, \ch{Br6HfPd}, \ch{Br6PdTi}, \ch{Br6Pt2}, \ch{BrCl5PdPt}, \ch{HfI6Pd}, \ch{BrCl5CuTi}, \ch{BrI5PtTi} &\parbox[c]{1em}{   \includegraphics[width=\smallpic]{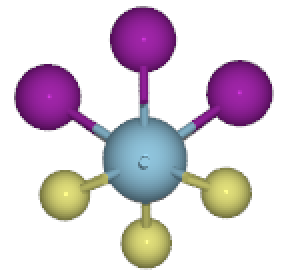}}&\parbox[c]{1em}{   \includegraphics[width=\smallpic]{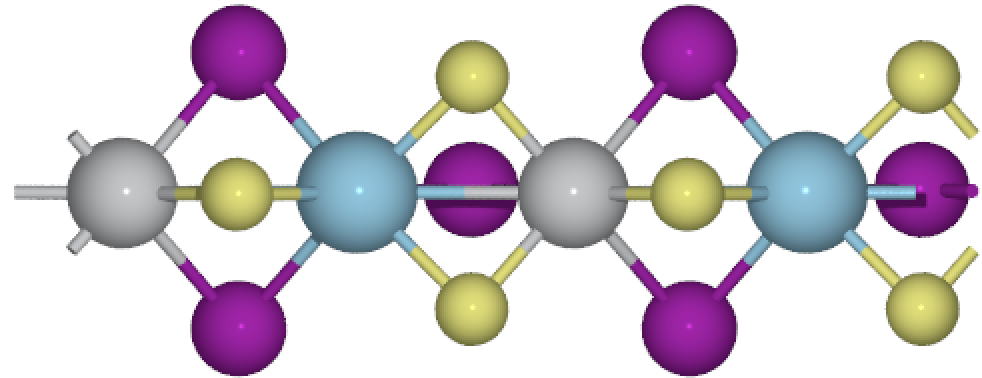}} \\ \hline

\textbf{Group 2}:\ch{I5Nb3Te2}, \ch{I5Nb3SeTe}, \ch{I4SeSiTe2TiZr}, \ch{I4Se3SiZr2}, \ch{Cl4HfRuS3Ti}, \ch{I4SeTe2TiZr2}, \ch{AlCl3S4Ti2}, \ch{Br3ClNbS3Ta2}, \ch{AlBr4CuS3Ti}, \ch{AlBrI4S2Ti2}, \ch{Br4ClNb3S2}, \ch{GeI5Se2VZr}, \ch{I4SnTe3TiZr}, \ch{Br4GeS3SiTa}, \ch{Br7GeMoNb}&\parbox[c]{1em}{   \includegraphics[width=\smallpic]{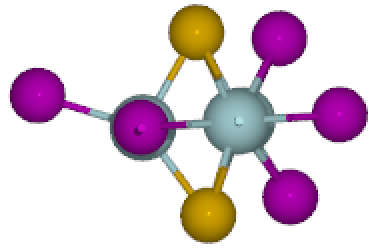}}&\parbox[c]{1em}{   \includegraphics[width=\smallpic]{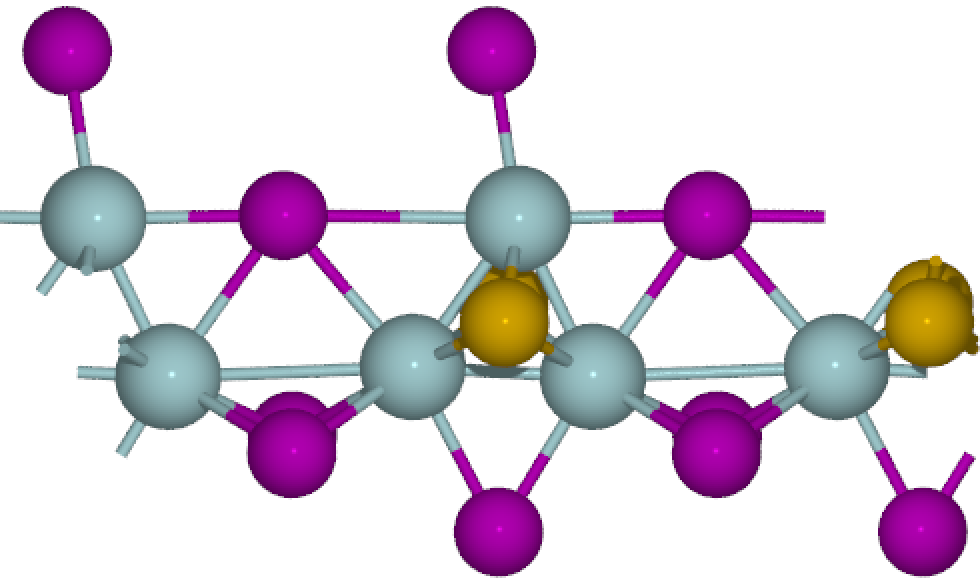}} \\ \hline

\textbf{Group 3}:\ch{Br8TiZr}, \ch{HfI8Zr}, \ch{Br5HfI3Zr}, \ch{Br8PtZr}, \ch{Cl8PtZr}, \ch{Br6Cl2PtTi}, \ch{Cl8PdTi}, \ch{Cl8PdPt}, \ch{BrI7Pt2}, \ch{Br7ClPtTa}, \ch{Cl8OsW}, \ch{Br3Cl5MoW}& \parbox[c]{1em}{
      \includegraphics[width=\smallpic]{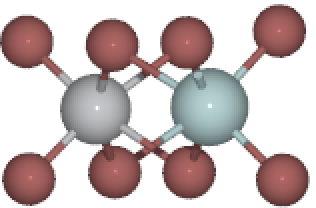}}& \parbox[c]{1em}{
      \includegraphics[width=\smallpic]{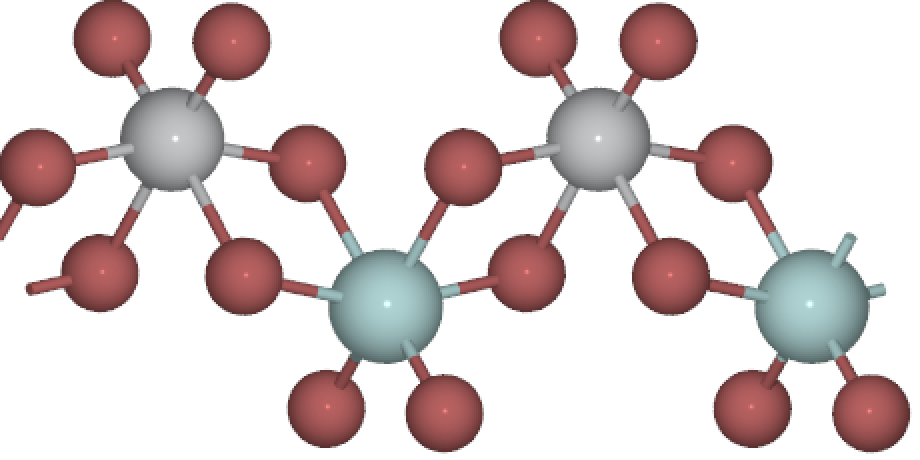}}\\ \hline

\textbf{Group 4}:\ch{Br3ClMo}, \ch{BrCl7W2}, \ch{C2I2W}, \ch{Br3ClPt}, \ch{Br4Pt}, \ch{BrCl7Pt2}, \ch{Cl8OsPt}, \ch{Br4I4PtW}, \ch{Cl8OsSn}, \ch{I3TaTe} & \parbox[c]{1em}{
      \includegraphics[width=\smallpic]{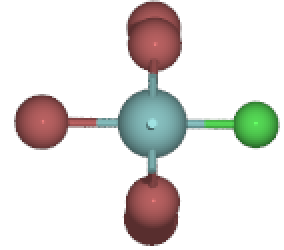}}& \parbox[c]{1em}{
      \includegraphics[width=\smallpic]{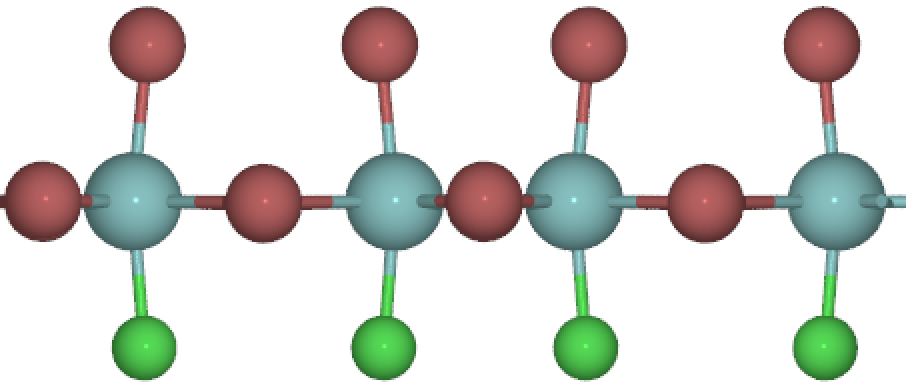}}\\ \hline

\textbf{Group 5}:\ch{Br8Hf4Se2}, \ch{Br8S2Ti4}, \ch{Hf4I8Se2}, \ch{Cl8SSeTi4}, \ch{Br8Hf4SeTe}, \ch{Br8Hf4Te2}, \ch{Cl8RuSe2Ti3}, \ch{Br2Cl8Hf2Zr2}, \ch{I8NbSeTa3Te}&\parbox[c]{1em}{
      \includegraphics[width=\smallpic]{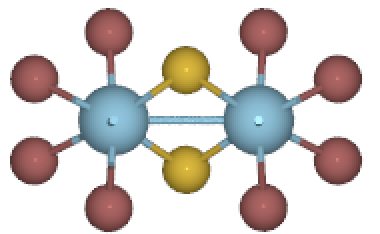}}& \parbox[c]{1em}{
      \includegraphics[width=\smallpic]{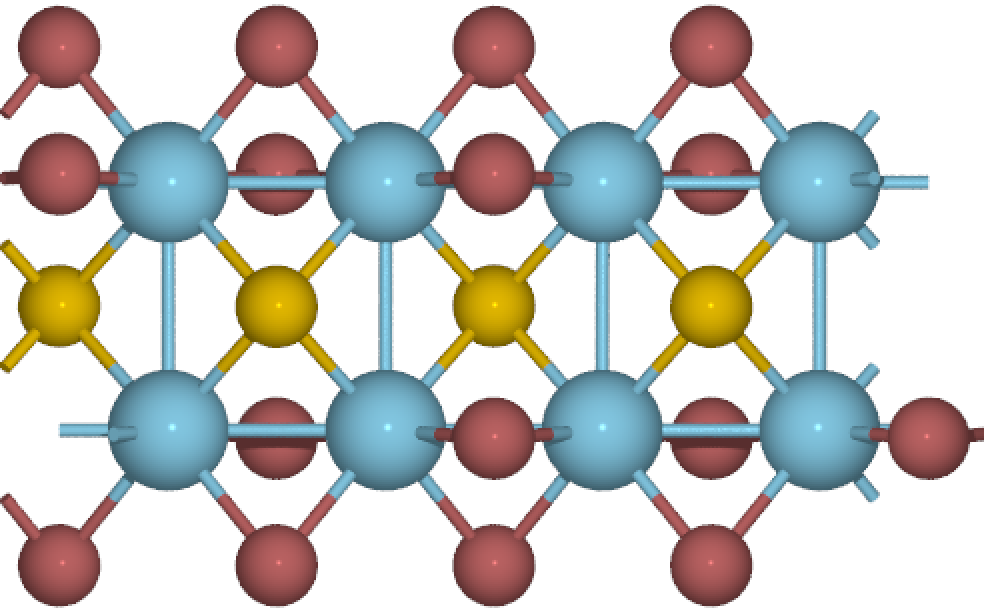}}\\ \hline

\textbf{Group 6}:\ch{I8SW3}, \ch{BrI7Mo3Se}, \ch{I8STa3}, \ch{BrI7STa3}, \ch{Br7ClNbSTa2}, \ch{Hf3I8S}, \ch{Cl8Mo2OW}, \ch{ClI6S2Ta3}&\parbox[c]{1em}{
      \includegraphics[width=\smallpic]{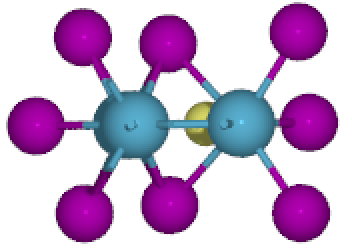}}& \parbox[c]{1em}{
      \includegraphics[width=\smallpic]{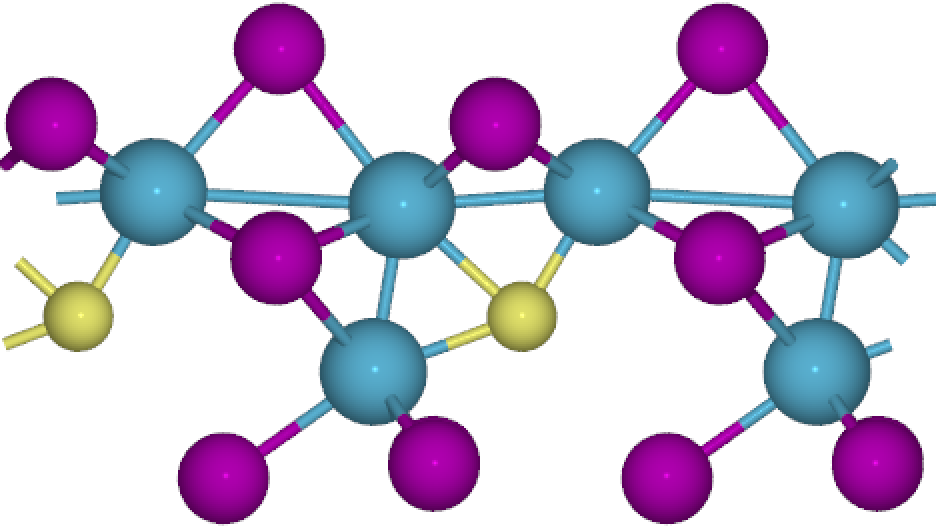}}\\ \hline
      
\textbf{Group 7}:\ch{I5MoS2Ti}, \ch{I5MoS2V}, \ch{Br5MoSe2V}, \ch{Br5MoSe2Ti}, \ch{Br5MoS2V}, \ch{Br5NbS2Ta}, \ch{Cl5MoS2W}&\parbox[c]{1em}{
      \includegraphics[width=\smallpic]{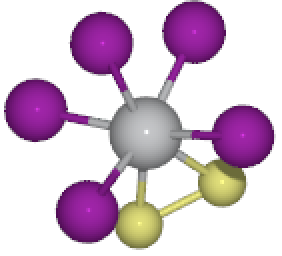}}& \parbox[c]{1em}{
      \includegraphics[width=\smallpic]{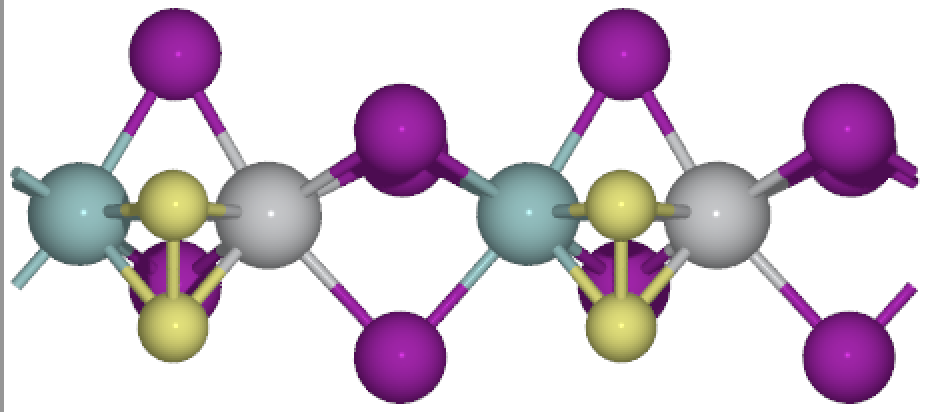}}\\ \hline

\textbf{Group 8}:\ch{Br5HfI3SeTiZr}, \ch{I8SeTi2V}, \ch{I7MoSe2Ti2}, \ch{Br7Nb3Te2}, \ch{Cl9PtW2}&\parbox[c]{1em}{
      \includegraphics[width=\smallpic]{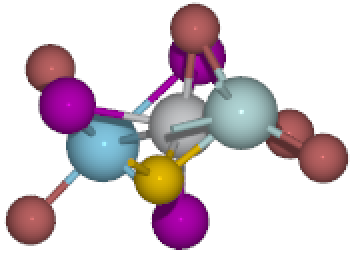}}& \parbox[c]{1em}{
      \includegraphics[width=\smallpic]{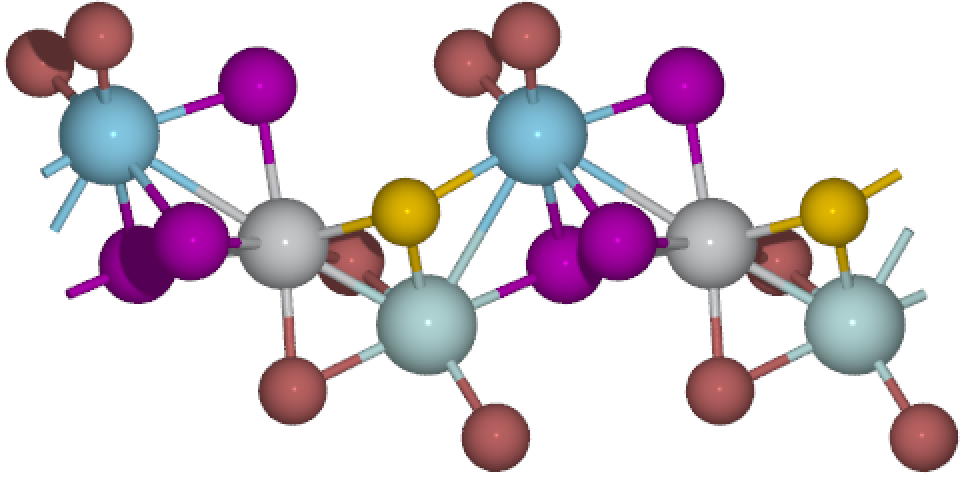}}\\ \hline

\textbf{Group 9}:\ch{Cl9HfMoTi}, \ch{Br9HfMoTa}, \ch{Cl8Mo2SSi}, \ch{BeCl8Mo2S}, \ch{AlCl9Ti2}&\parbox[c]{1em}{
      \includegraphics[width=\smallpic]{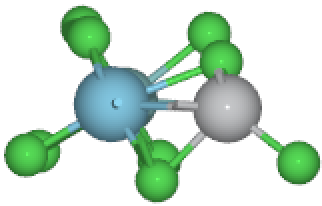}}& \parbox[c]{1em}{
      \includegraphics[width=\smallpic]{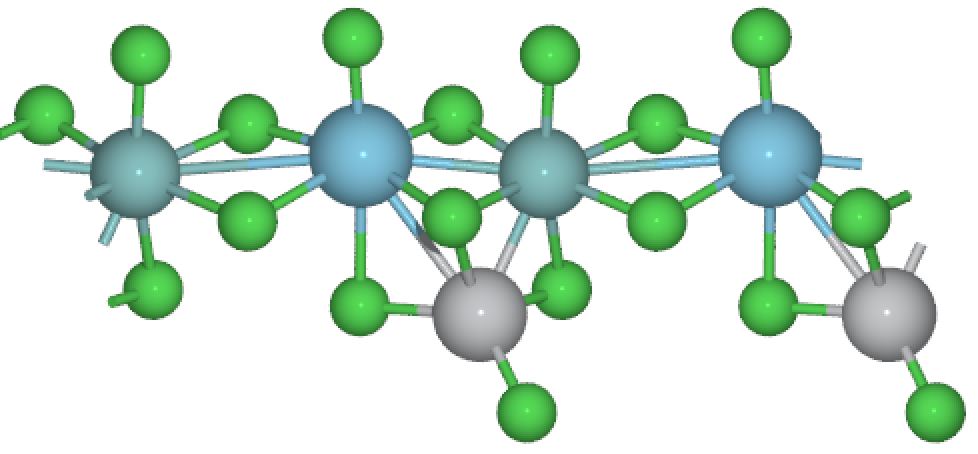}}\\ \hline
      
\textbf{Group 10}:\ch{Br8ClHfPtSn}, \ch{BiBr7Hf2Se2}, \ch{Cl9OsPtSn}, \ch{Br2I4Pt3Se3}, \ch{Cl9TiW2}&\parbox[c]{1em}{
      \includegraphics[width=\smallpic]{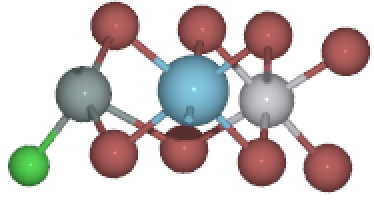}}& \parbox[c]{1em}{
      \includegraphics[width=\smallpic]{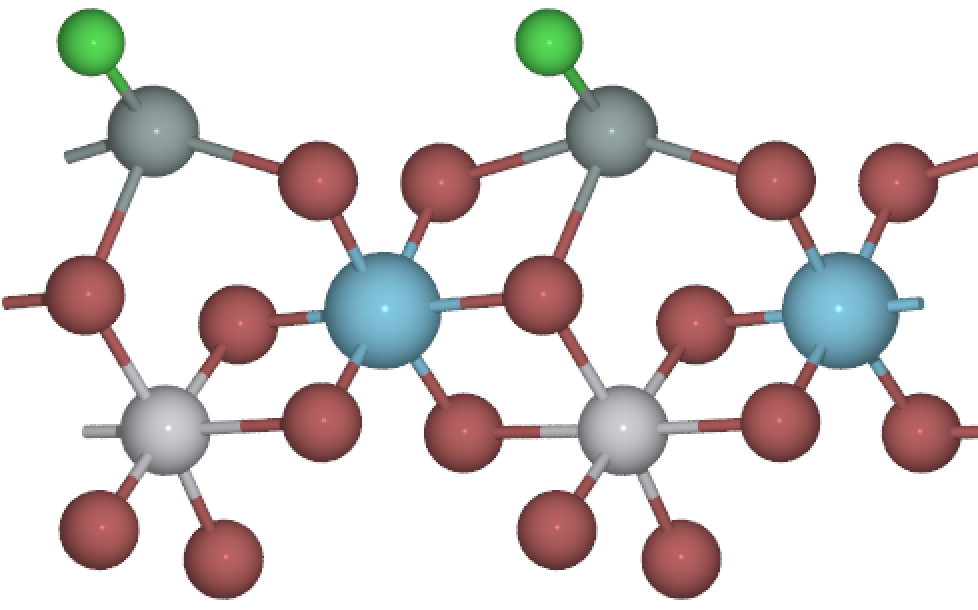}}\\ \hline
\end{tabular}
\caption{Clusters of crystal structures discovered using the dendrogram. The table displays all clusters with more than five materials.}\label{tab:structure_type}
\end{table} 

It turns out that ten of the classes contain five or more materials and they are listed in Table~\ref{tab:structure_type}. 

For several of the classes, there is an obvious connection between the class and the position of the elements in the periodic table. For example, for the largest group (Group 1) with 34 materials the materials typically have two transition metal atoms and six halogen atoms. In some cases the halogen atoms are substituted by chalcogen atoms. Around a row of transition metal atoms, the halogens/chalcogens position themselves with an approximate threefold symmetry. There are also some exceptions as for example \ch{Bi2I6} and \ch{Bi2Br6}, where \ch{Bi} substitutes the transition metal atoms. There is of course some variation in bond distances and symmetry, but the overall structure of the compounds remain the same.

Group 3 seems to be a generalization of Group 1 but now with two transition metal atoms and eight halogen atoms per unit cell. The transition metal atoms now are not directly bound to each other, but the binding is mediated by the extra halogen atoms.

Group 5 is also similar to group 1 but with two chains of transition metal atoms and with four halogen atoms and one chalcogen atom per two transition metal atoms. The transition metal atoms are again surrounded by halogen and chalcogen atoms with an approximate threefold symmetry, where the the halogen atoms are bound to a single transition-metal atom, while the divalent chalcogen atoms bonding to two transition-metal atoms thereby connecting the two chains.

Group 7 can also be derived from Group 1, but with an interesting substitution where one of the halogen or chalcogen atoms is replaced by a dimer of chalcogen atoms. Since the dimer is expected to be divalent it can substitute a single chalcogen atom.

It is interesting to analyze in which ways the machine-generated structures differ from already know structures from C1DB. As mentioned above C1DB consists of two parts, the core, which is obtained by exfoliation of experimentally known structures, and the shell, which is generated by element substitution in the core systems. The element substitutions are performed in such a way that all atoms of a given element are substituted at the same time, i.e. if three sulfur atoms appear in a certain material, then all three atoms may be substituted by, say, selenium, but only at the same time. This also means that symmetries, where for example three sulfur atoms are rotated onto each other, will be conserved in the materials with the substitutions. 

The machine clearly generates new materials by a more general substitution of chemical elements. The largest group of materials in this study (Group 1) clearly resemble the largest group of materials in the core of C1DB (named Group 1 in Ref.~\onlinecite{moustafa2022computational}), which contains materials like \ch{ZrI3}, \ch{TiCl3}, and \ch{MoBr3}. These materials also consist of chains of transition-metal atoms surrounded by halogen atoms in a triangular pattern, but because the halogen atoms are identical the three-fold rotational symmetry is perfect. The Group 1 materials generated by the machine can be obtained by element substitution in the core materials from C1DB, but where individual atoms of the same chemical element can now be substituted by different atoms thereby breaking some symmetries. However, by using the RMSD as a structural distance measure the similarity of the structures appears independent of symmetry breaking.

We have investigated the connection between the geometrically classified groups in Table~\ref{tab:structure_type} and the groups in the core of C1DB more systematically by again applying RMSD similarity with single-linkage clustering and a cutoff of $R_0 = 0.75 \textrm{\AA}$ as used in the previous work. The analysis shows that Group 1 is in fact close to the Group 1 of Ref.~\onlinecite{moustafa2022computational}. Furthermore Group 3 is similar to Group 6 for the core materials, and Group 4 is similar to Group 4 of Ref.~\onlinecite{moustafa2022computational}. Groups 3 and 4 could therefore possibly have been obtained by a more general chemical element substitution than the one performed in the previous work. However, some of the other groups, like Group 5, do not have any similar materials in C1DB as an indication that the machine is also able to generate completely new structures and not only perform element substitution. Some of the new groups might be obtained by substituting single atoms by more complex radicals, but Group 5 is an example of a new and highly symmetric class not seen before. 

\subsection{Band structure}

\begin{figure}
\centering
\includegraphics[width=0.8\linewidth]{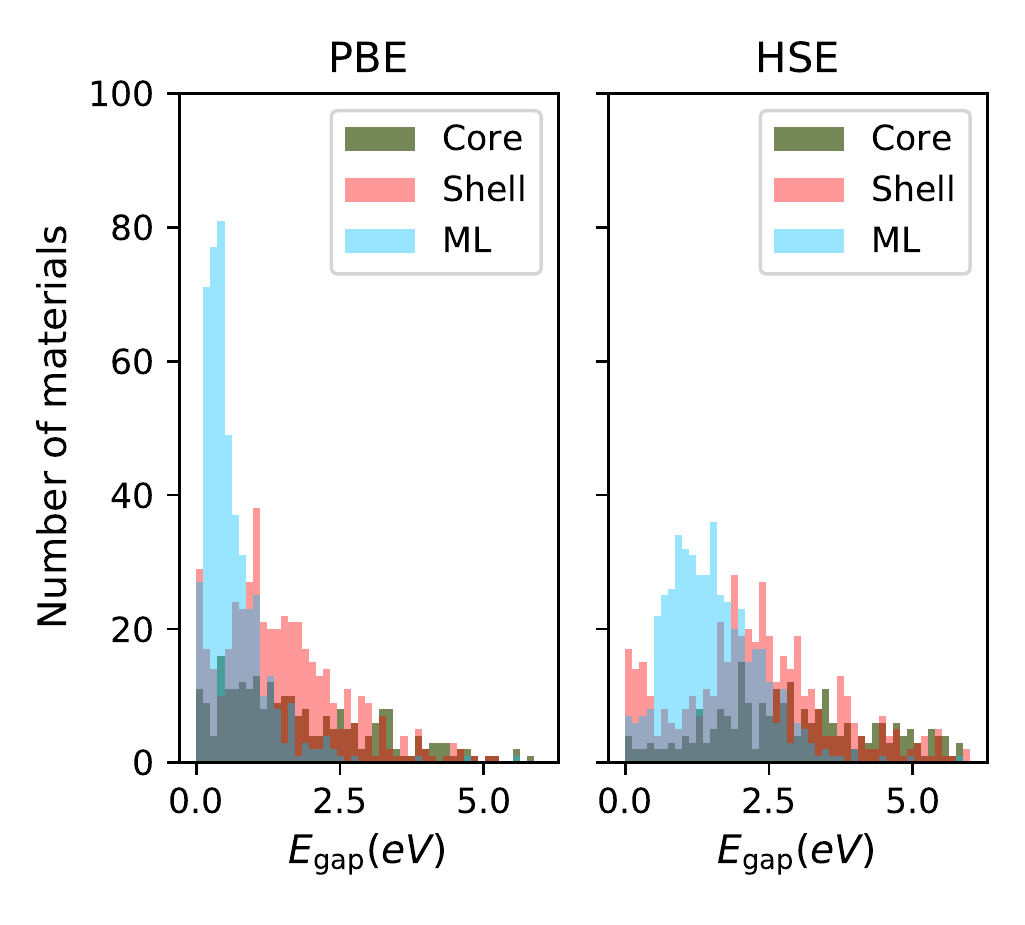}
\caption{The energy gaps distribution as determined by PBE and HSE}
\label{fig:band_gap_PBE_and_HSE}
\end{figure}

We have calculated the band structures of all the new materials with both PBE and non-selfconsistent HSE06. These are all available in the new addition to C1DB \cite{c1db}, so here we shall only show some statistics for the calculated band gaps. Figure \ref{fig:band_gap_PBE_and_HSE} displays the distribution of band gaps for the one-dimensional components as calculated by PBE and HSE06. Data are shown for both the core and shell of C1DB and for the new machine-generated materials. Only the materials with non-vanishing band gaps are shown in the figure. Among the new materials, there are 46 (40) metallic systems according to the PBE (HSE06) calculations.

Fig~\ref{fig:band_gap_PBE_and_HSE} shows the expected shift to larger band gaps when comparing HSE06 to PBE. PBE is known to generally underestimate the band gaps, while HSE06 provides values closer to experiment. Interestingly, the machine-generated materials are seen to have on the average lower band gaps than the ones in the core and shell of C1DB. This results in many new materials with band gaps in the visible range of the electromagnetic spectrum. Some of the new materials might therefore be relevant as light-absorbers.

\subsection{Effective masses}
\begin{figure}
     \centering
         \includegraphics[width=0.9\columnwidth]{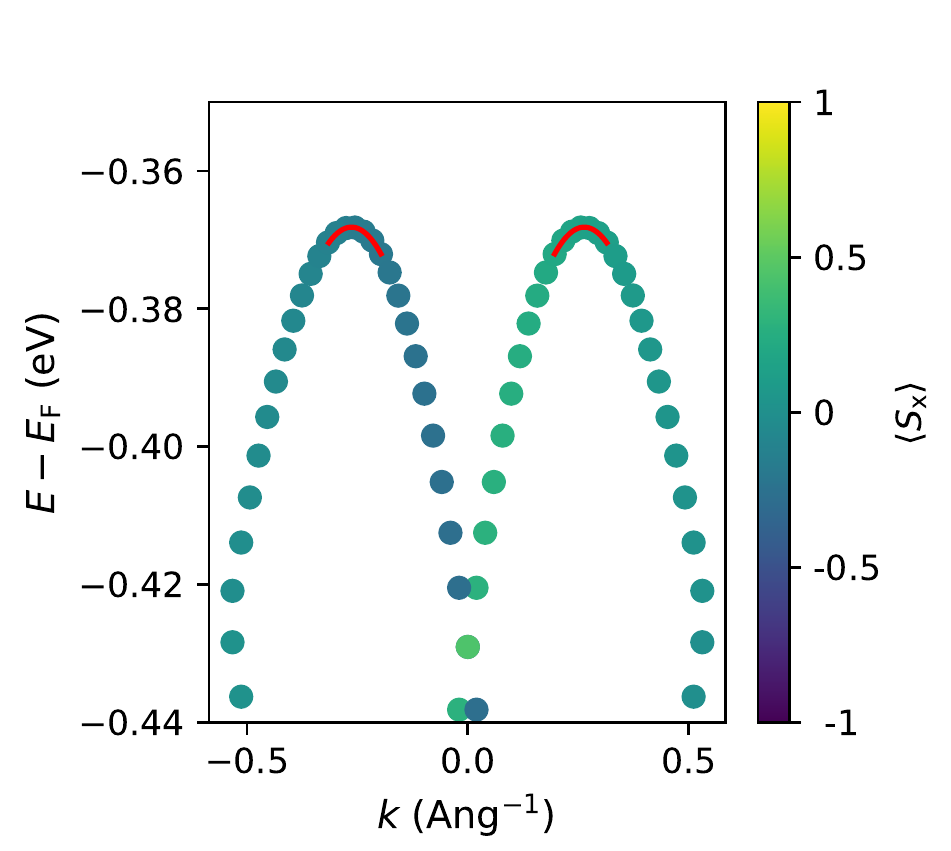}
        \caption{The top of the valence band in \ch{Cl8Pt2S}. The colors of the circles indicate the size of the x-component of the spin. The hole mass is determined by a parabolic fit as illustrated in the figure. A large spin-orbit splitting is clearly observed.}
        \label{fig:masses}
\end{figure}
The effective electron and hole masses, $m^\ast_{e,h}$ are calculated for all systems with a band gap using parabolic fits as described in the methods section. Again, all calculated data are available in the new version of C1DB. Fig.~\ref{fig:masses} illustrates in the case of the system \ch{Cl8Pt2S} the upper part of the valence band and the determination of the hole mass. The band structure is seen to exhibit a significant Rashba splitting.

As discussed in Ref.~\onlinecite{moustafa2022computational} atomically thin one-dimensional systems with large Rashba splittings may potentially exhibit Majorana bound states. We therefore also calculate the wave vector $k_R$ at the band edge and the Rashba parameter $\alpha_R = \hbar^2k_R/(2m^\ast)$, and include them in the database. For holes in \ch{Cl8Pt2S} shown in Fig.~\ref{fig:masses}, the values are $k_\mathrm{R}=0.26$ \AA$^{-1}$,
$m^\ast =4.4\, m_e$, and $\alpha =0.46$ eV \AA.

\section{Conclusions}
The machine-learning method applied in this work is shown to be quite efficient in generating new, stable materials. Out of the 1895 unique, new one-dimensional materials suggested by the generator, 529 materials fulfill the two stability criteria of less than 0.2 eV above the convex hull and positive phonon frequencies. 

A part of the new materials can be viewed as generalizations of the materials already present in C1DB, where the core materials come from experimental databases, and where the shell is constructed by element substitution. In C1DB no materials have more than four different elements, while no such limitation exists for the new materials, which have up to 8 different elements. Some of the new materials have clearly the same basic atomic structure as materials already present in C1DB, but with more general substitution of similar elements.

However, the CDVAE also generates completely new classes of materials with structures not seen before. Some of these structures are even geometrically rather simple with an understandable chemistry. A nice example of this is Group 5 in Table~\ref{tab:structure_type} where two transition metal chains decorated by halogen atoms are combined through the substitution of two halogen atoms by a bonding divalent chalcogen atom.

It is an open question to which extent the new materials will be experimentally synthesizable either in bulk form with weakly bound one-dimensional components or as individual one-dimensional chains, which could potentially be formed at a substrate. According to the calculations, many of the new materials would have the sufficient thermodynamic and dynamical stability to exist, but the calculations give no indication of realistic synthesis paths. Additional information about entropic effects can be obtained through further analysis of the phonon spectra, but it is an outstanding challenge to computationally address possible synthesis procedures.

The new materials are added to the C1DB \cite{c1db} database with the keyword Source set to 'Machine learning generated'. This keyword distinguishes them from materials in the core (Source = 'COD' or 'ICSD') and the shell (Source = 'Derived by element substitution') of the database.

\section{Acknowledgements}
The VILLUM Center for Science of Sustainable Fuels and Chemicals, which is supported by the VILLUM Fonden research grant 9455, is acknowledged by K.W.J. and H.M.
Additionally, we thank financial support from the European Research Council (ERC) provided via the Horizon 2020 research and innovation program of the European Union under Grant No. 773122 (LIMA) and Grant agreement No. 951786. (NOMAD CoE). K.S.T. receives support from VILLUM FONDEN as a Villum Investigator (grant no. 37789).

\bibliography{refs}
\end{document}